\documentclass[prd,twocolumn,superscriptaddress,showpacs,preprintnumbers,amsmath,amssymb,APS]{revtex4}
\usepackage{bm}
\usepackage[latin1]{inputenc}
\usepackage[spanish,english]{babel}
\usepackage{amsfonts}
\usepackage{amssymb}
\usepackage{graphicx}
\usepackage{color}

\newcommand{\be}{\begin{equation}}
\newcommand{\ee}{\end{equation}}
\newcommand{\bea}{\begin{eqnarray}}
\newcommand{\eea}{\end{eqnarray}}

\begin{document}

\title{{\bf Conformal anomaly and primordial magnetic fields}}

\author{Ivan Agullo}\affiliation{\footnotesize  Department of Physics and Astronomy, Louisiana State University, Baton Rouge, LA 70803-4001;}\affiliation{\footnotesize CTC, DAMTP, University of Cambridge, Wilberforce Road, Cambridge, CB3 OWA, U.K.}
\author{José Navarro-Salas}
\affiliation{ {\footnotesize Departamento de Fisica Teorica and
IFIC, Centro Mixto Universidad de Valencia-CSIC,   Facultad de Fisica, Universidad de Valencia,
        Burjassot-46100, Valencia, Spain.}}

\date{September 20, 2013}

\begin{abstract}

The conformal symmetry of the quantized electromagnetic field breaks down in curved space-time. We point out that this conformal anomaly is able to generate a sizable magnetic field during a phase of slow-roll inflation. Such primordial magnetism is characterized by the  expectation value of the squared of the magnetic field  for comoving observers, which at leading order in slow-roll takes the value $\langle\vec B^2\rangle =\frac{8}{15(4\pi)^2}\,  H^4\epsilon$, where $\epsilon$ is the standard slow-roll parameter. This result is insensitive to the intrinsic ambiguities of  renormalization in curved space-times. The information in the quantum state gets diluted during inflation and does not affect the prediction. A primordial  coherent field  of this strength may be able to seed the observed cosmic magnetism.

\end{abstract}


\maketitle

{\it Introduction.} A long standing open question in cosmology concerns the origin of observed coherent ${\mu}$G magnetic fields in galaxies and galaxy clusters (see \cite{kandus, Subramanian, widrowetal} for reviews). Although different types of astronomical dynamos have been proposed, they only provide amplification mechanisms which need of pre-existing magnetic fields, with intensities between $10^{-12}$ and $10^{-22}$ G, to operate. In addition,  difficulties in accounting for the magnetism observed in high-redshift protogalaxies and  in empty intergalactic space (the latter with strength  $\sim 10^{-17}-10^{-14}$ G)  \cite{tavecchio}  remain.

An attractive possibility is that primordial magnetic fields originated during inflation via the same mechanism that produces the scalar perturbations responsible for the temperature anisotropies in the cosmic microwave background. As pointed out in \cite{turner-widrow}, inflation  provides ideal conditions to account for an ubiquitous primordial magnetism at large-scales. There is however a significant difficulty with this proposal \cite{turner-widrow}. The Maxwell electromagnetic Lagrangian  $\mathcal{L}^{M}= -1/4\sqrt{-g}F_{\mu\nu}F^{\mu\nu}$, where $F_{\mu\nu}= \nabla_{\mu} A_{\nu} - \nabla_{\nu} A_{\mu}$ and is $A_{\mu}$ the vector potential, is conformal invariant (unlike a minimally coupled scalar field). Since the space-time metric during inflation is conformally flat,  the quantum mechanism responsible of amplifying scalar inflaton perturbations is not present for the electromagnetic field.  Consequently, although inflation naturally generates  large-scale magnetic fields, it is believed that their strength is
 rapidly diluted by the exponential expansion (typically as $\vec{B}^2\sim a^{-4}$) and it is unable to account for observations. To bypass this conclusion one needs to assume new physics that explicitly break the conformal electromagnetic invariance \cite{turner-widrow}, and predictions become model dependent.

In this note we exploit the fact that the conformal symmetry of the Maxwell Lagrangian is spontaneously broken in the quantum theory, without the need of introducing new ingredients in the theory. This is the well known conformal or trace anomaly which arises in quantum electrodynamics in {\it curved} space-times, {\it  even for the free Maxwell Lagrangian} $\mathcal{L}^{M}$ (it appears also for other conformally invariant fields). The conformal anomaly has played an important role in understanding the existence and properties of Hawking radiation by black holes \cite{christensen-Fulling, conformal-kerr, fabbri-navarro}, and we argue here that it can also account for a primordial magnetic field  with the desired properties generated during {\it slow-roll} inflation. Our conventions follow \cite{MTW}, with $\hbar =1 = c$.

{\it Renormalization in curved space-times and conformal anomaly.} 
The intensity of the magnetism generated in the early universe  can be quantified by the expectation value $\langle \vec{B}^2\rangle$, where $ \vec{B}$ is the magnetic field as measured by a comoving cosmological observer.  This expectation value is ultraviolet divergent, even in the non-interacting theory, and renormalization and regularization techniques are required to extract the physically relevant result. 
Renormalization techniques in curved space-times were extensively studied in the 70's \cite{birrell-davies, Waldbook, parker-toms}. The best studied object is the energy-momentum tensor $T_{\mu\nu}$, which plays an important role in the theory. Our goal is to extract the value of $\langle \vec{B}^2\rangle$ from the well known results for the renormalized  $\langle T_{\mu\nu}\rangle_{\rm ren}$.

To quantize the electromagnetic field one has to deal with gauge issues. A convenient way of proceed in curved space-times is by introducing a term in the Lagrangian  to fix the gauge, e.g. the covariant Lorentz gauge in which $\nabla^{\mu}A_{\mu}=0$, and then include a complex ghost field $c$ to restore the gauge invariance (see e.g. \cite{birrell-davies}). The formal energy-momentum tensor is then given by 
$\langle  T_{\mu\nu} \rangle =   \langle  T^M_{\mu\nu} \rangle +  \langle  T^{BR}_{\mu\nu} \rangle +  \langle  T^{GH}_{\mu\nu} \rangle $. It was shown in  \cite{Adler77} that the  ghost and the gauge-breaking contributions exactly cancel each other,  $\langle  T^{BR}_{\mu\nu} \rangle +  \langle  T^{GH}_{\mu\nu}\rangle=0$, and the non-trivial result arises entirely from the expectation value of the classical Maxwell energy-momentum tensor $T_{\mu\nu}^{M} = -\frac{1}{4}g_{\mu\nu} F^2 + F_{\mu\alpha}F_{\nu}^{\ \alpha}$, {\it together with} geometric contributions coming from renormalization (see below). The conformal invariance of the classical theory is reflected in the vanishing trace, $g^{\mu\nu}T^{M}_{\mu\nu}=0$.

 The most general approach to renormalize $\langle  T_{\mu\nu} \rangle$ was developed by Wald \cite{wald78, Waldbook}, who proposed a set of natural physical criteria, or axioms, that any candidate $\langle T_{\mu\nu}\rangle_{\rm ren}$ to the right-hand-side of the Einstein equations should satisfy. The remarkable consequence is that this set of axioms characterize the value of $\langle T_{\mu\nu}\rangle_{\rm ren}$ {\it almost} uniquely, up to a conserved, local curvature tensor \cite{wald78, Waldbook}. For the electromagnetic field the result is \cite{Adler77, brown-ottewill86} (see also \cite{wald78} for the scalar field)
\be \label{Tmn}  \langle  T_{\mu\nu} \rangle_{ren} =   \langle  T^{M (B)}_{\mu\nu} \rangle +  \mathcal{T}_{\mu\nu} + c_1\, I_{\mu\nu}+c_2 \, J_{\mu\nu}\ . \ee
In this expression $\langle  T^{M(B)}_{\mu\nu} \rangle \equiv \langle -\frac{1}{4}g_{\mu\nu} F^2 + F_{\mu\alpha}F_{\nu}^{\ \alpha} \rangle_{ren}$ is a {\it traceless, but not conserved} tensor \footnote{Defined in this way, $\langle  T^{M(B)}_{\mu\nu} \rangle=\lim_{x' \to x} \tau_{\mu\nu\rho\sigma}[W_{\alpha}^{\rho\sigma}]$, where $\tau_{\mu\nu\rho\sigma}$ is a differential operator in terms of which the classical energy-momentum tensor can be written as $T^M_{\mu\nu}=\lim_{x' \to x}\tau_{\mu\nu\rho\sigma}[A^{\rho}(x)A^{\sigma}(x')]$, and $W_{\alpha}^{\rho\sigma}(x,x')$ is the state dependent, smooth (non-divergent) part of the Hadamard expansion of the vector potential two-point function for the state $|\alpha\rangle$ (see e.g. \cite{brown-ottewill86}). One could argue that there exist the freedom of including an arbitrary, geometric, traceless and not conserved tensor into the definition of $\langle  T^{M(B)}_{\mu\nu} \rangle$, along with the subtraction of the same tensor from $\mathcal{T}_{\mu\nu}$. The above choice  is however preferred, since it  modifies as little as possible the classical expression $T^M_{\mu\nu}$, in the spirit of minimal subtraction. Additionally, note that the traceless tensor $\tau_{\mu\nu\rho\sigma}[W_{\alpha}^{\rho\sigma}]$ is the only term contributing to the {\it difference} of expectation values $\langle \alpha |  T_{\mu\nu}| \alpha \rangle_{ren}-\langle \beta |  T_{\mu\nu}| \beta \rangle_{ren}=\lim_{x' \to x} \tau_{\mu\nu\rho\sigma}[W_{\alpha}^{\rho\sigma}-W_{\beta}^{\rho\sigma}]$.}. Following the notation in \cite{Adler77}, the subscript $B$ refers to `boundary-condition-dependent', indicating that  it encodes all the dependence on the quantum state;  $\mathcal{T}_{\mu\nu}$ is the local, geometric tensor 
\be \mathcal{T}_{\mu\nu}= \frac{-1}{2(4\pi)^{2}}\left[(-\frac{3}{4}a_{2 \ \rho}^{\rho}+a_2)g_{\mu\nu} +a_{2 \mu\nu}\right] \, , \ee
where $a_{2 \mu\nu}$ and $a_2$ are the  DeWitt coefficients associated with the vectorial and scalar  wave equations (see e.g \cite{birrell-davies}). It is such that $\nabla^{\mu}(\langle  T^{M(B)}_{\mu\nu} \rangle+\mathcal{T}_{\mu\nu})=0$; $I_{\mu\nu}$ and $J_{\mu\nu}$ are the two independent, conserved tensors with the same dimensions as $T_{\mu\nu}$, constructed out from the local geometry, which can be obtained, respectively,  by functional variation of the Lagrangian densities $\sqrt{-g}R^2$ and $\sqrt{-g} R_{\mu\nu}R^{\mu\nu}$ with respect to the metric tensor. They have a  trace $I^{\mu}_{\mu}=3J^{\mu}_{\mu}=-6\Box R$, where  $R_{\mu\nu}$ is the Ricci tensor and $R$ its trace \cite{wald78, Waldbook}; the arbitrary real numbers $c_1$ and $c_2$ parametrize the ambiguity in the renormalization procedure.  

An important implication of the principles of  renormalization in curved space-time \cite{wald78, birrell-davies, Waldbook, parker-toms} is the unavoidable existence of a non-vanishing trace $\langle  T \rangle \equiv g^{\mu\nu} \langle T_{\mu\nu}\rangle_{\rm ren}=\mathcal{T}^{\mu}_{\mu} -2 (3c_1+c_2) \Box R$, as first noticed in \cite{capper-duff}
\be \label{tracegeneric}\langle  T \rangle  = \frac{1}{(4\pi)^2} \left(\frac{1}{10} C_{\mu\nu\alpha \beta}C^{\mu\nu\alpha \beta}-\frac{31}{180}G+ C\ \Box R\right)\, , \ee
where $G= R_{\mu\nu\alpha \beta}R^{\mu\nu\alpha \beta}-4 R_{\mu\nu}R^{\mu\nu}+ R^2$ is the Gauss-Bonnet topological invariant, with  $C_{\mu\nu\alpha \beta}$, $R_{\mu\nu\alpha \beta}$ the Weyl and the Riemann tensor, respectively. $C=-1/10-2 \, (3\, c_1+c_2)$ (note that in de Sitter space $R$ is constant and therefore $\Box R=0$: the renormalization ambiguity {\it completely disappears}). This trace is a quantum effect,  is {\it independent} of the state in which the  expectation value is evaluated, and is constructed from local geometric tensors. This is the celebrated conformal or  trace anomaly, which constitutes a robust prediction of renormalization in curved space-times.  Note that the existence of this anomaly is not manifest in the equations of motion. They are still conformal invariant. Therefore, in the case of Friedman-Lemaitre-Robertson-Walker (FLRW) space-times discussed below,   the non-zero trace is consistent with the absence of particle (photons) creation \cite{Parker79, parker-toms}. The conformal anomaly is a collective effect which can not be understood by looking at a single Fourier mode; it rather arises as a consequence of the locality and covariance of quantum field theory in curved space-times \cite{hollands-wald}.

In the presence of interactions with other fields, additional contributions to the trace (\ref{tracegeneric}) may appear \cite{birrell-davies}. Those terms are generically non-geometrical and may depend on the quantum state. They are however suppressed by the coupling constants,  and generally provide sub-leading corrections to the free trace anomaly.

{\it The renormalized stress-energy tensor in a FLRW space-time}.
Our next goal is to obtain $\langle  T^{M(B)}_{\mu\nu} \rangle$ in terms of the conformal anomaly. It will suffice for our purposes to focus on the time-time component, although the rest can be easily obtained from it by symmetry arguments.  We will take advantage from a result due to Parker \cite{Parker79} (see also \cite{birrell-davies, parker-toms}): in the FLRW space-time 
with line element $ds^2=g_{\mu\nu}dx^{\mu}dx^{\nu}=-dt^2+a^2(t)d\vec{x}^2$, the quantity $ \langle T_{00}\rangle_{ren} \equiv t^{\mu}t^{\nu} \langle T_{\mu \nu}\rangle_{ren}  $, where $t^{\mu}=(\partial/\partial t)^{\mu}$, evaluated in a translational invariant state can be completely determined from $\langle T \rangle $, up to an integration constant $\alpha$ related to the quantum state. The proof relies on the existence of a conformal Killing vector field, i.e. a vector field $k^{\mu}$ such that $\mathsterling_{k}g_{\mu\nu}\equiv2 \nabla_{(\mu} k_{\nu)}= \gamma(x) \, g_{\mu\nu}$. For the FLRW metric, $k^{\mu}= a(t) \, t^{\mu}$ is a conformal Killing vector  with $\gamma=2\dot a(t)$ (dot indicates derivative with respect to $t$). Using this, and the fact that $\langle T_{\mu\nu}\rangle_{ren}$ is a symmetric and conserved tensor, we have 
\be \nabla^{\mu}( \langle T_{\mu\nu}\rangle_{ren} \, a\, t^{\nu})= \dot a \, \langle T \rangle \, . \ee
One can now integrate this expression over the four-volume bounded by constant time hyper-surfaces $t_1$ and $t_2$. By using the divergence theorem and assuming the quantum state is translational invariant,  we arrive at

\be \nonumber a(t_2)^4  \langle T_{00}(t_2)\rangle_{ren} -a(t_1)^4  \langle T_{00}(t_1)\rangle_{ren}=-\int_{t_1}^{t_2} dt\, a^3 \dot a\,  \langle T \rangle\ee
In Minkowski space-time $k^{\mu}$ becomes a Killing vector field and this equation provides the conservation of energy. In FLRW, on the contrary, the energy density dilutes as $a^{-4}$. Additionally, if a non-vanishing trace that breaks conformal invariance is present, the energy density gets `sourced' by it. We find
\be \label{B22}  \langle T_{00}\rangle_{ren}=\frac{1}{a^4} (-g(t)+\alpha), \ee
where $g(t)$ is determined by the trace anomaly via $\partial_{t} g(t)=a^3\dot a \langle T \rangle $, with $\alpha$ an integration constant  encoding the information of the state in which the expectation value is evaluated.  The function $g(t)$ admits the local solution 
\be \nonumber \label{g} g(t)= \frac{6}{(4\pi)^4}[\frac{-31}{180}\dot a^4 +C(-a^2\dot a \dddot a - a\dot a^2\ddot a +\frac{1}{2}a^2\ddot a^2+\frac{3}{2}\dot a^4)] \ . \ee
The constant $\alpha$ in (\ref{B22}) allows for the possibility that the state contains an initially homogenous distribution of photons, with an energy density that is redshifted as $a^{-4}$. The remaining terms, which are the relevant ones for our considerations,  can be thought of as due to virtual pairs or vacuum polarization \cite{Parker79}. 

We can now take advantage of the above results to get an expression for the quantity $\langle  T^{M(B)}_{\mu\nu} \rangle t^{\mu}t^{\nu}= 1/2\langle \vec E^2  +\vec B^2 \rangle$ ($\vec{E}$ and $\vec{B}$ are the electric and magnetic fields defined by a comoving cosmological observer. Equivalently, we could have focused on any of the other diagonal components  of $\langle  T^{M(B)}_{\mu\nu} \rangle$; the analysis would be analogous and the final result the same). Using equations (\ref{Tmn}) and (\ref{B22}), we get

\be \label{TM}\langle  T^{M(B)}_{\mu\nu} \rangle t^{\mu}t^{\nu} = \frac{1}{a^4} (-g(t)+\alpha) -  \mathcal{T}_{\mu\nu}t^{\mu}t^{\nu} - (c_1+c_2/3)I_{\mu\nu} t^{\mu}t^{\nu}\  \ee
where we have taken into account that in FLRW the two conserved geometric tensors $I_{\mu\nu}$ and $J_{\mu\nu}$ are not independent, $I_{\mu\nu}= 3J_{\mu\nu}$. An important property of equation (\ref{TM}) is that, although the renormalization ambiguity enters in the first and last terms of the right hand side (note that $g(t)$ depends on $c_1$ and $c_2$), both contributions {\it exactly cancel out}, making the left hand side {\it independent of the choice of renormalization procedure}. This is a major simplification, which can be easily understood by noticing that in FRLW the  only conserved geometric tensor with the appropriate dimensions, $I_{\mu\nu}$, has non-zero trace, and therefore can not contribute to the traceless part of the renormalized  energy-momentum tensor.  

Equation (\ref{TM}) takes the form
\bea \label{E2B2}\frac{1}{2}\langle \vec E^2 + \vec B^2 \rangle&=&   \frac{1}{240(4\pi)^2a^4}[ 194\dot a^4 -242 a\dot a^2 \ddot a +48a^2 \ddot a^2  \nonumber \\ &-&6a^2\dot a \dddot a 
+ 6a^3\ddddot a] + \frac{\alpha}{a^4}\ , \eea
where we have made use of the identity 
\bea \label{t00} \mathcal{T}_{\mu\nu}t^{\mu}t^{\nu}
&=& \frac{1}{240 (4\pi)^2 a^4} (270 {\dot a}^4+98a \dot a^2 \ddot a+ 24 a^2 \ddot a^2 \nonumber \\ &-&138 a^2\dot a \dddot a -6 a^3\ddddot a) 
\ . \eea  
Note that the term containing the information of the initial states in (\ref{E2B2}) dilutes as $a^{-4}$.

For an exact de Sitter expansion, where $a(t)=e^{Ht}$, a remarkable simplification occurs in  (\ref{E2B2}):
\be \label{B2desitter}  \frac{1}{2}(\langle \vec E^2 \rangle  + \langle \vec B^2 \rangle)=  \frac{\alpha}{a^4}  \ . \ee
Furthermore, for the de Sitter invariant state we find  $\alpha=0$. This result is  a consequence of the underlying symmetries, which make the expectation value of the energy-momentum in the de Sitter invariant state to be proportional to the metric tensor, enforcing any traceless part to vanish. Notice also  that the energy density does not vanish, $\langle T_{00}\rangle_{ren}=\frac{31}{16(4\pi)^2} H^4$, although it produces negligible back-reaction when $H$ is the inflationary Hubble rate (this is also true in slow-roll inflation).

{\it Electromagnetic duality.}
To extract  an expression for  $\langle \vec{B}^2 \rangle$ from equation (\ref{E2B2})  we will take advantage of the invariance of the electromagnetic equations of motion under the duality transformation 
$F^{\mu\nu} \to {^{*}F}^{\mu\nu}= \frac{1}{2}\epsilon^{\mu\nu\lambda\sigma}F_{\lambda\sigma}$, which translates into the same relation for the corresponding  quantum operators. Given a quantum state,  this relation would imply $\langle F_{\mu\nu}F_{\lambda\sigma}\rangle = \langle  {^{*}F}_{\mu\nu}{^{*}F}_{\lambda\sigma}\rangle$, and therefore $ \langle \vec E^2  \rangle= \langle \vec B^2 \rangle$, only in case the state  is also  invariant \cite{Adler77}. However, for arbitrary states one can still  ensure  equality of the {\it state-independent} information in $\langle \vec E^2  \rangle$ and $ \langle \vec B^2 \rangle$. Therefore, equation (\ref{E2B2}) gives
\bea \label{B2} \langle \vec B^2 \rangle&=&   \frac{1}{240(4\pi)^2a^4}[ 194\dot a^4 -242 a\dot a^2 \ddot a +48a^2 \ddot a^2  \nonumber \\ &-&6a^2\dot a \dddot a 
+ 6a^3\ddddot a] + \frac{\alpha_B}{a^4}\ , \eea
with  a similar expression holding for $\langle \vec E^2 \rangle$ with $\alpha_B \to \alpha_E$, where $\alpha_B$ and $\alpha_E$ are constants obtained from the quantum state, satisfying $\frac{1}{2}(\alpha_B + \alpha_E)= \alpha$. Furthermore, the state-dependent term is washed away by the expansion at the rate  $a^{-4}$. Consequently, during inflation the magnetic field quickly forgets about the details of the quantum state, unlike the case of scalar inflaton perturbation where the process of stimulated creation of quanta compensates for the dilution \cite{agullo-parker}.

{\it Primordial magnetic fields.}
In an exact de Sitter expansion $a(t)=e^{Ht}$, as already pointed out, a cancelation of the state-independent terms takes place and  $\langle \vec B^2 \rangle=  \frac{\alpha_B}{a^4}$.  
Notice that, had we included a non vanishing curvature $K$ for the spatial sections, an extra term proportional to $K H^2/a^2$ would appeared, which is also diluted by the expansion, this time at  the rate $a^{-2}$ rather than $a^{-4}$ (see also the analysis of \cite{barrow}). 

 However, in a realistic model of inflation the Hubble rate $H(t)\equiv \dot a/a$ is time-dependent and the previous simplification is not present. 
Working at leading order in the slow-roll approximation, we have $ a\dot a^2 \ddot a = \dot a ^4(1-\epsilon)$, $a^2\ddot a^2 = \dot a ^4(1-2\epsilon)$, 
$a^2 \dot a \dddot a = \dot a ^4(1-3\epsilon)$, $a^3 \ddddot a = \dot a ^4(1-6\epsilon)$, where $\epsilon$ is the slow-roll parameter $\epsilon \equiv -\dot H/H^2 \ll 1$.
The terms in (\ref{B2}) sum up to 
\be \label{res} \langle \vec B^2 \rangle =  \frac{8}{15(4\pi)^2}H^4\epsilon + \mathcal{O}^2\ , \ee
where $\mathcal{O}^2$ encodes terms of quadratic or higher order in slow-roll (we have dropped out the state-dependent term, which is diluted  with the expansion). This constitutes our main result. Note that this result also applies for  a more fundamental $U(1)$ gauge field such as the hypercharge, which relates to the electromagnetic field via the Weinberg angle, $A_{\mu}=Y_{\mu} \cos \theta_W$.

With the assumed  range for the inflationary energy scale, of order  $10^{15}-10^{16}$GeV, $\epsilon\sim 10^{-2}$,  and using standard arguments of entropy conservation (for simplicity we also assume that the Universe transited to radiation domination immediately after the end of inflation \cite{Subramanian}) one gets an order of magnitude for the primordial magnetic field of $B_0 \sim 10^{-16}-10^{-14}$G, which suffices to seed the galactic dynamo and fits well with the strength of the observed magnetic field in intergalactic space.

A somewhat related computation has been considered in \cite{campanelli} in  exact de Sitter space-time using adiabatic regularization, along the lines of the analysis  of the scalar power spectrum performed in references \cite{parker07}. Our results differ from those in \cite{campanelli}, where $\langle \vec B^2 \rangle$ is reported to be different from zero for an exact  de Sitter expansion (notice that dimensional arguments enforce $\langle \vec B^2 \rangle$ to be proportional to $H^4$). Adiabatic regularization is not manifestly covariant, since its applicability is restricted to homogenous space-times. One therefore needs to proceed with care in other to reproduce results that are in accordance with the axiomatic theory of renormalization, particularly for massless, conformal invariant fields (see e.g. \cite{bunch80, chimento-cossarini}).

{\it Conclusions.}
 We have shown that the conformal anomaly of the electromagnetic field
together with symmetry arguments, lead to a primordial magnetic field statistically homogeneous and isotropic, of size given by equation (\ref{res}), generated in slow-roll inflation. The result is  independent of the choice of renormalization procedure, and of the quantum state describing the gauge field at the onset of inflation. Our approach is specially suited to rigorously determine the amplitude $\langle \vec B^2 \rangle$ of the coherent magnetic field, and makes the gravitational conformal anomaly a good candidate to explain the  strength of the ubiquitous magnetism observed in the cosmos.
 
 {\it Acknowledgments:}   This work is supported  by the  Spanish Grants FIS2011-29813-C02-02,  CPANPHY-
1205388, the European project MP1210-MoU.CSO,  Hearne Institute and EU Marie Curie funds.  I.A. thanks C. Bonvin, L. Pogosian, and specially A. Ashtekar for stimulating discussions.

\end{document}